\documentclass[preprints,article,accept,moreauthors,pdftex]{Definitions/mdpi} 

\firstpage{1} 
\pubvolume{xx}
\issuenum{1}
\articlenumber{5}
\pubyear{2019}
\copyrightyear{2019}
\history{Received: date; Accepted: date; Published: date}

\pdfoutput=1

\newcommand\figref[1]{Figure~\ref{#1}}

\Title{Role of Coherent Eddies in Potential Vorticity Transport in Two-layer Quasigeostrophic Turbulence}

\Author{Wenda Zhang $^{1,*}$, Christopher L.P. Wolfe $^{1}$ and Ryan Abernathey $^{2,}$}

\AuthorNames{Wenda Zhang, Christopher L.P. Wolfe and Ryan Abernathey}

\address{%
$^{1}$ \quad School of Marine and Atmospheric Sciences, Stony Brook University, Stony Brook, NY 11794, USA; wenda.zhang@stonybrook.edu; christopher.wolfe@stonybrook.edu\\
$^{2}$ \quad Lamont-Doherty Earth Observatory, Columbia University, New York, USA; rpa@ldeo.columbia.edu}

\corres{Correspondence: wenda.zhang@stonybrook.edu} 

\abstract{The transport by materially coherent eddies is studied in a two-layer quasigeostrophic model of geophysical turbulence.  The coherent eddies are identified by closed contours of the Lagrangian-averaged vorticity deviation \cite{Haller_Hadjighasem_2016} obtained from Lagrangian particles advected by the flow.  A series of flow regimes with different bottom friction strengths are considered---it is found that coherent eddies become more prevalent and longer-lasting as the strength of the bottom drag increases.  These coherent eddies, with average core radius close to the deformation radius, propagate zonally with speeds close to the long baroclinic Rossby wave speed and meridionally with a preference for cyclones to propagate poleward and anticyclones to propagate equatorward.  The meridional propagation preference of the coherent eddies gives rise to a systematic upgradient potential vorticity (PV) transport, which is in the opposite direction as the background PV transport and not captured by standard Lagrangian diffusivity estimates.  The upgradient PV transport by coherent eddy cores is less than 10\% of the total PV transport, but the PV transport by the periphery flow induced by the PV inside coherent eddies is significant and downgradient.  This clarifies the distinct roles of the trapping and stirring effect of coherent eddies in PV transport in geophysical turbulence.}

\keyword{ocean mesoscale eddies; Lagrangian coherent structure, geostrophic turbulence; PV transport; eddy diffusivity}

\begin{document}

\section{Introduction}
Ocean mesoscale eddies play an important role in the transport and mixing in the ocean, and can affect wide range of oceanographic phenomena, including the stratification and overturning circulation in the Southern Ocean \cite{Hallberg_Gnanadesikan_2006, Marshall_Radko_2003, Marshall_Radko_2006, Wolfe_Cessi_2009,Wolfe_Cessi_2010}, eastern boundary currents \cite{Griffiths_Pearce_1985, Fang_Morrow_2003, Kurian_Colas_2011, Cessi_Wolfe_2009, Colas_Capet_2013, Pelland_Eriksen_2013, Bire_Wolfe_2018, Steinberg_Pelland_2019}, nutrient cycling in the upper ocean \cite{McGillicuddy_2003}, and ocean uptake of carbon dioxide \cite{Siegenthaler_1983, Gnanadesikan_2015}.  In the classical view, an ``eddy'' is typically defined as a deviation from a suitably defined mean state and is thought to contribute to tracer transport primarily by stirring background tracer gradients.  However, with the advent of satellite observations and high-resolution ocean models, the observed eddies are increasingly considered as individual coherent structures \cite{Chelton_2011, Zhang_Wang_2014, Zhang_Zhang_2017}.  Coherent eddies can trap fluid inside their cores and physically move water and tracers over potentially long distances.  The importance of this transport mechanism to ocean circulation is currently uncertain.  This study investigates the contribution of coherent eddies to tracer transport in a two-layer quasigeostrophic model of geophysical turbulence.

Coherent eddies (vortices) are a characteristic phenomenon in geophysical turbulence.  Freely decaying barotropic turbulence tends to organize itself into a finite number of isolated coherent vortices starting from random initial conditions \cite{McWilliams_1984}.  These vortices are distinct objects with closed circulations which maintain their identity for times much longer than eddy turnover time \cite{Provenzale_1999}.  Although the formation of these coherent vortices is believed to be related to the inverse energy cascade, the specific formation mechanisms are still uncertain.  Studies have shown that coherent vortices are important in determining the mixing and dispersion properties of turbulent flows \cite{Provenzale_1999, Bracco_LaCasce_2000}.  On the one hand, the strong PV gradients around coherent vortices work as material barriers---impermeable to fluid inside vortex cores---so fluids or passive tracers are trapped inside the vortices for times comparable to the vortex lifetime.  This leads to nonlocal and subdiffusive mixing \cite{Provenzale_1999, Berloff_McWilliams_2002a}.  On the other hand, in flows characterized as ``vortex-dominated'', Lagrangian particle dispersion \cite{Pasquero_Provenzale_2001} and Eulerian eddy fluxes \cite{Thompson_Young_2006} outside the vortex cores can also be dominated by the vortex dynamics.  Despite their recognized importance, the flow regimes that favor the formation of coherent vortices and their transport characteristics remain open questions.

The majority of modern ocean models---including the oceanic components of climate models---are of insufficient resolution to resolve mesoscale eddies. The effects of eddies must therefore be parameterized in terms of large-scale quantities that are resolved by the model.  The eddy flux is typically represented as a diffusive process where the flux of a tracer is related to the large-scale tracer gradient by an eddy diffusivity.  However, a diffusive parameterization does not fully account for the effects of coherent eddies \cite{Provenzale_1999, Pasquero_Provenzale_2001}.  \citet{Bracco_LaCasce_2000} showed that the existence of coherent vortices leads to non-Gaussian velocity distributions, and \citet{Berloff_McWilliams_2002a} revealed the nondiffusive properties of the tracer transport by mesoscale eddies in oceanic gyres, which casts doubt on using horizontally isotropic eddy diffusion for eddy parameterization.  Studies by \citet{Pasquero_Provenzale_2001} and \citet{Berloff_McWilliams_2002b} further showed that parameterizations using stochastic models to account for effects of coherent vortices can better simulate the turbulent dispersion.  It is therefore necessary to understand the dynamics and transport properties of coherent eddies in order to appropriately parameterize their effects.

Estimates of the transport by coherent eddies in ocean are quite uncertain.  A factor which contribute to this uncertainty is the lack of a widely-accepted and rigorous definition of a ``coherent vortex''.  \citet{Chelton_2011} assumed eddies exceeding a threshold value of a nonlinearity parameter were coherent and found that most of the mesoscale motions observable from satellite were coherent.  \citet{Zhang_Wang_2014} defined a ``coherent eddy'' as a closed potential vorticity (PV) contour and estimated that the zonal volume transport due to these coherent eddies was 30--40 Sv in the subtropics.  Another study by \citet{Dong_McWilliams_2014} detected coherent eddies based on the geometry of the velocity vectors and showed that the meridional heat and salt transport due to the individual coherent eddies is significant.  However, the eddies in these studies were detected by Eulerian methods that can be shown to be non-objective; that is, observers in different reference frames disagree about the boundary or even the existence of the coherent eddy \cite{Haller_2005}. 

The particle advection studies of \citet{Beronvera_Wang_2013}, \citet{Haller_Beronvera_2013}, and \citet{Wang_Olascoaga_2015} have shown that eddies identified by Eulerian methods are very leaky---the defined boundary does not effectively trap the fluid inside it.  \citet{Haller_2005} developed an objective definition of a coherent eddy as a rotationally-coherent Lagrangian vortex (RCLV) and, in \citet{Haller_Hadjighasem_2016}, showed that the cores of RCLVs are maxima of the Lagrangian averaged vorticity deviation (LAVD)---vorticity averaged along the trajectories of Lagrangian particles.  The LAVD method reliably detects RCLVs and is significantly less computationally expensive than the previous objective coherent eddy detection methods \cite{Hadjighasem_Farazmand_2017}.  Applying this method, \citet{Abernathey_Haller_2018} showed that the RCLVs in eastern Pacific are generally smaller and shorter-lived than the eddies detected by \citet{Chelton_2011}.  As a result, the zonal volume transport by these RCLVs was estimated to be one order smaller than \citet{Zhang_Wang_2014}'s estimate, and their meridional transport was negligible.  Thus, the previous estimates using Eulerian methods generally overestimated the importance of coherent eddies, which causes the uncertainty about their role in tracer transport.  This uncertainty extends to the fundamental question of whether coherent eddy transport is a significant player in oceanic tracer transport.

The aim of this study is to quantify the contribution of coherent eddies to tracer transport in the classical setting of two-layer quasigeostrophic (QG) turbulence.  The advantage of the QG system is its simplicity, generality, and controllability which allows us to consider a range of oceanographically relevant regimes and obtain results which are reasonably independent of the vagaries of individual observational and modeling systems.  Following \citet{Abernathey_Haller_2018}, we apply the LAVD method to detect coherent eddies.  Instead of estimating the volume flux due to coherent eddies we focus on the PV flux, which is a crucial dynamical quantity in geophysical flows.  PV is a materially conservative tracer, and its flux can affect the evolution of the flow.  The meridional PV flux---the sum of the heat and vorticity flux---plays an important role in both driving the mean flow and maintaining the eddy enstrophy budget.  Quantifying the contribution of coherent eddies to the meridional PV transport sheds light on the dynamical role of coherent eddies in the eddy-mean flow interaction.  The QG model we used here mimics the dynamics in Southern Ocean, and several different values of bottom drag are used to simulate different flow regimes.

This paper is structured as follows.  In section \ref{method}, we introduce the two-layer QG model, the Lagrangian diagnostics and the detection method of coherent eddies.  Section \ref{eddy_statistics_sec} provides the statistics of coherent eddies in a series of different friction cases with implications for their transport properties.  In section \ref{sec_PV_trans}, we present the meridional PV transport due to coherent eddies and their contribution to the total PV transport.  In section \ref{diff_sec}, we estimate the Lagrangian diffusivity due to coherent eddies, which quantifies the transport of generic passive tracers.  Section \ref{discussion_sec} is the discussion and conclusions.

\section{Methods}

\label{method}
\subsection{Model description}

This study uses the Python package \texttt{pyqg} \cite{Abernathey_Rocha_2016} to simulate two-layer QG turbulence.  The governing equations of the model are the forced-dissipative PV evolution equations in the two layers:
\begin{linenomath}
\begin{align}
\label{eq:qg_layer1}
\frac{\partial q_{1}}{\partial t} + U_{1}\frac{\partial q_{1}}{\partial x} + J(\psi_1,q_1) + [F_1(U_1 - U_2) + \beta]\frac{\partial \psi_1}{\partial x} &= \mathrm{ssd},\\
\label{eq:qg_layer2}
\frac{\partial q_{2}}{\partial t} + U_{2}\frac{\partial q_{2}}{\partial x} + J(\psi_2,q_2) + [F_2(U_2 - U_1) + \beta]\frac{\partial \psi_2}{\partial x}&= -r_{ek}\nabla^2\psi_2+ \mathrm{ssd},
\end{align}
\end{linenomath}
where the PV in layer $n (n=1,2)$ is
\begin{linenomath}
\begin{equation}
\label{eq_PV}
q_n = \nabla^2 \psi_n + (-1)^n F_n(\psi_1 - \psi_2).
\end{equation}
\end{linenomath}
The flow is forced by a background vertical shear, $\Delta U  =  U_1 - U_2$.  The Jacobian is
\begin{linenomath}
\begin{equation}
J(\psi,q) = \frac{\partial \psi}{\partial x}\frac{\partial q}{\partial y} - \frac{\partial \psi}{\partial y}\frac{\partial q}{\partial x}
\end{equation}
\end{linenomath}
and
\begin{linenomath}
\begin{align}
F_1&=\frac{k_d^2}{1+\delta}, &  F_2 &=\frac{k_d^2\delta}{1+\delta},
\end{align}
\end{linenomath}
where $k_d$ is the inverse of the Rossby deformation radius, $L_d$, and $\delta = H_1/H_2$ is the ratio of the thickness of two layers.  

We use the same parameters as \citet{Wang_Jansen_2016}: $L=1200$ km, $L_d=15$ km, $H_1=800$ m, $\delta=0.25$, $U_1=0.04$ m/s, $U_2=0$ and $\beta=1.3 \times 10^{-11}$ $\mathrm{m}^{-1}s^{-1}$.

The model is run in a doubly periodic domain with a horizontal resolution of $512\times512$ grid points.  A series of different values of $r_{ek}$ spanning an order of magnitude are used by this study.  Following \citet{Wang_Jansen_2016}, $r_{ek}^{-1}=10$, $20$, and $40$ days are taken as representative values of the drag parameter and referred to as the ``strong friction'', ``control'' and ``weak friction'' cases, respectively.  

\begin{figure}[H]
\centering
\includegraphics[width=\textwidth]{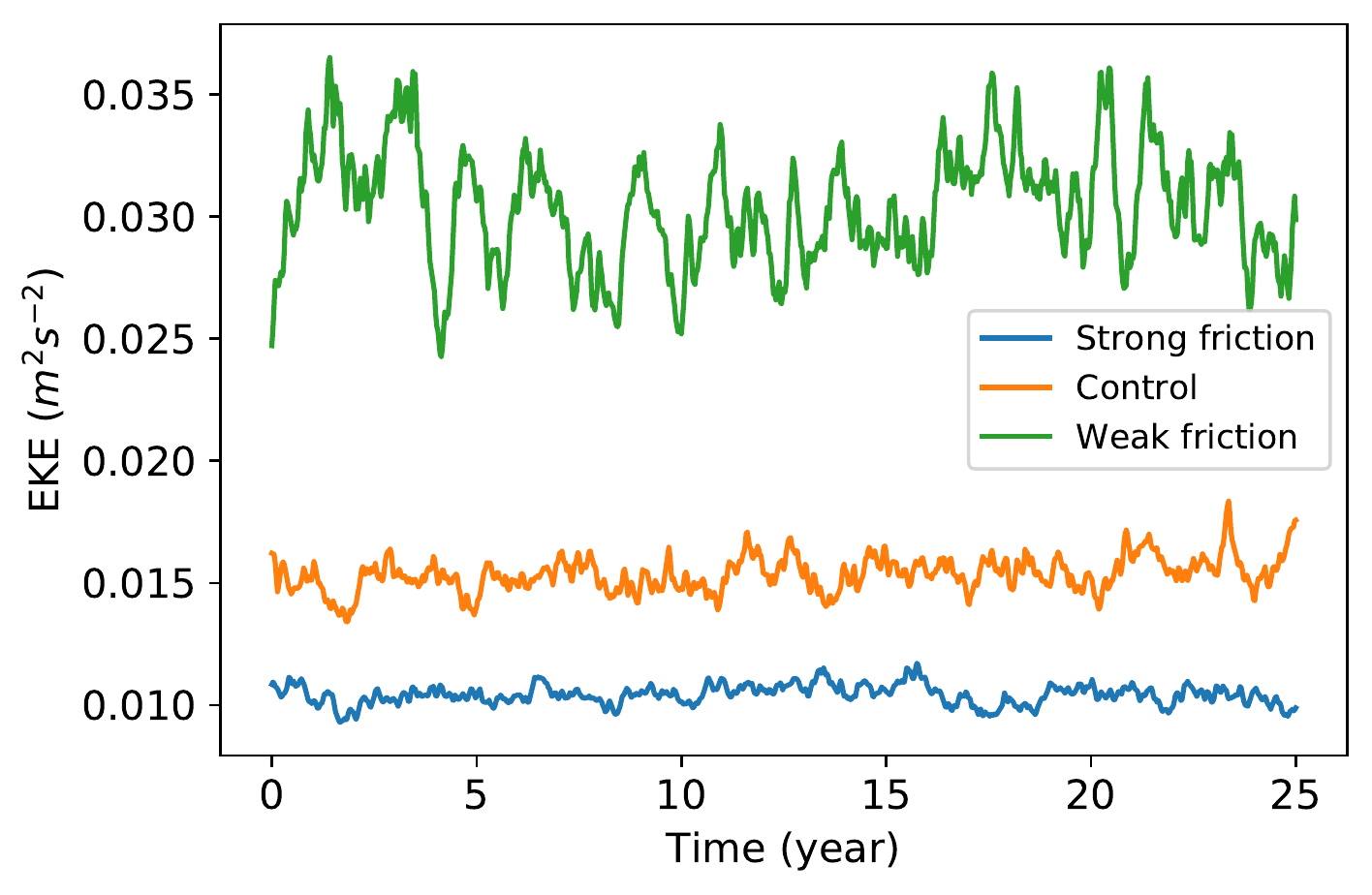}
\caption{Time series of upper layer EKE of the three simulations.  Blue, orange and green curves indicate the weak friction, control and strong friction cases, respectively.}
\label{eke_series}
\end{figure}

Each case was run for 60 years with time step of 0.3125 hours, and the last 50 years were sampled daily to calculate the diagnostics.  Figure \ref{eke_series} gives the eddy kinetic energy (EKE) time series in the upper layer over 25 years; here ``eddy'' refers to a deviation from the time-independent, zonal-mean flow.  The standard deviation of the EKE is 4\%\textendash8\% of its mean value, and all three experiments show episodes where the EKE grows or decays by 20\%\textendash30\%.  This implies that relatively long time averages will be necessary to obtain robust statistics.

\subsection{Lagrangian particle advection and Lagrangian diffusivity}

Once the flow is equilibrated, the upper layer is seeded with Lagrangian particles whose positions, $\mathrm{\mathbf{x}}$, evolve according to:
\begin{linenomath}
\begin{equation}
\frac{d\mathrm{\mathbf{x}}}{dt} = \mathrm{\mathbf{u}}[\mathrm{\mathbf{x}}(t),t]
\end{equation}
\end{linenomath}
where $\mathrm{\mathbf{u}}$ is the velocity.  The particles are initialized on a rectangular grid with a uniform spacing of 1.2 km for a total of 1,048,576 particles.  They are advected for 90 (for RCLV detection) or 360 (to estimate Taylor diffusivity) days with their positions, velocities, vorticities and PV saved daily.  The RCLV detection process was repeated for one hundred continuous and nonoverlapping intervals.

Particle trajectories are used to calculate the Lagrangian diffusivity.  According to \citet{Taylor_1922}, for homogeneous and stationary turbulent flow, the Eulerian diffusivity is consistent with the single-particle Lagrangian diffusivity which can be estimated as
\begin{linenomath}
\begin{equation}
\label{K_Lagr}
K(y_0,t)=\frac{1}{2}\frac{d}{dt}\left[\frac{1}{N}\sum_{i=0}^{N-1}(y_i(t) - y_{i0})^2\right]
\end{equation}
\end{linenomath}
where $y_i (t)$ is the meridional position of a particle released at $y_{i0}$ at time $t=0$ and $N$ is the total number of particles.

\subsection{Identification of coherent eddies}
\label{eddy_detection}
We use the Lagrangian Averaged Vorticity Deviation (LAVD) technique of \citet{Haller_Hadjighasem_2016} to detect coherent eddies, which are defined as a nested family of fluid tubes which rotate collectively around a common core.  The cores of coherent eddies are identified as the maxima of LAVD, defined as
\begin{linenomath}
\begin{equation}
\label{LAVD}
\mathrm{LAVD}_{t_0}^{t_1} (\mathrm{\mathbf{x}}_0) = \frac{1}{t_1 - t_0} \int_{t_0}^{t_1} |\zeta[\mathrm{\mathbf{x}}(\mathrm{\mathbf{x}}_0,t),t] - \overline{\zeta}|dt
\end{equation}
\end{linenomath}
where $\zeta[\mathrm{\mathbf{x}}(\mathrm{\mathbf{x}}_0,t),t]$ is the vorticity of the particle located at position $\mathbf{x}_0$ at time $t_0$, and $\overline{\zeta}$ is the domain-averaged vorticity (zero in the present case).  The integrand is the vorticity deviation---the absolute difference of vorticity from its domain average---so the LAVD field is the vorticity deviation averaged along the Lagrangian trajectories over the time interval $(t_0,t_1)$.  Closed contours of LAVD surrounding a maximum thus represent rings of particles rotating around the core of an eddy at the same average rate over the specified time interval.  The outmost convex LAVD contour is defined as the material boundary of the eddy which is expected to trap all the fluid parcels inside during that time interval. 

\begin{figure}[H]
\includegraphics[width=\textwidth]{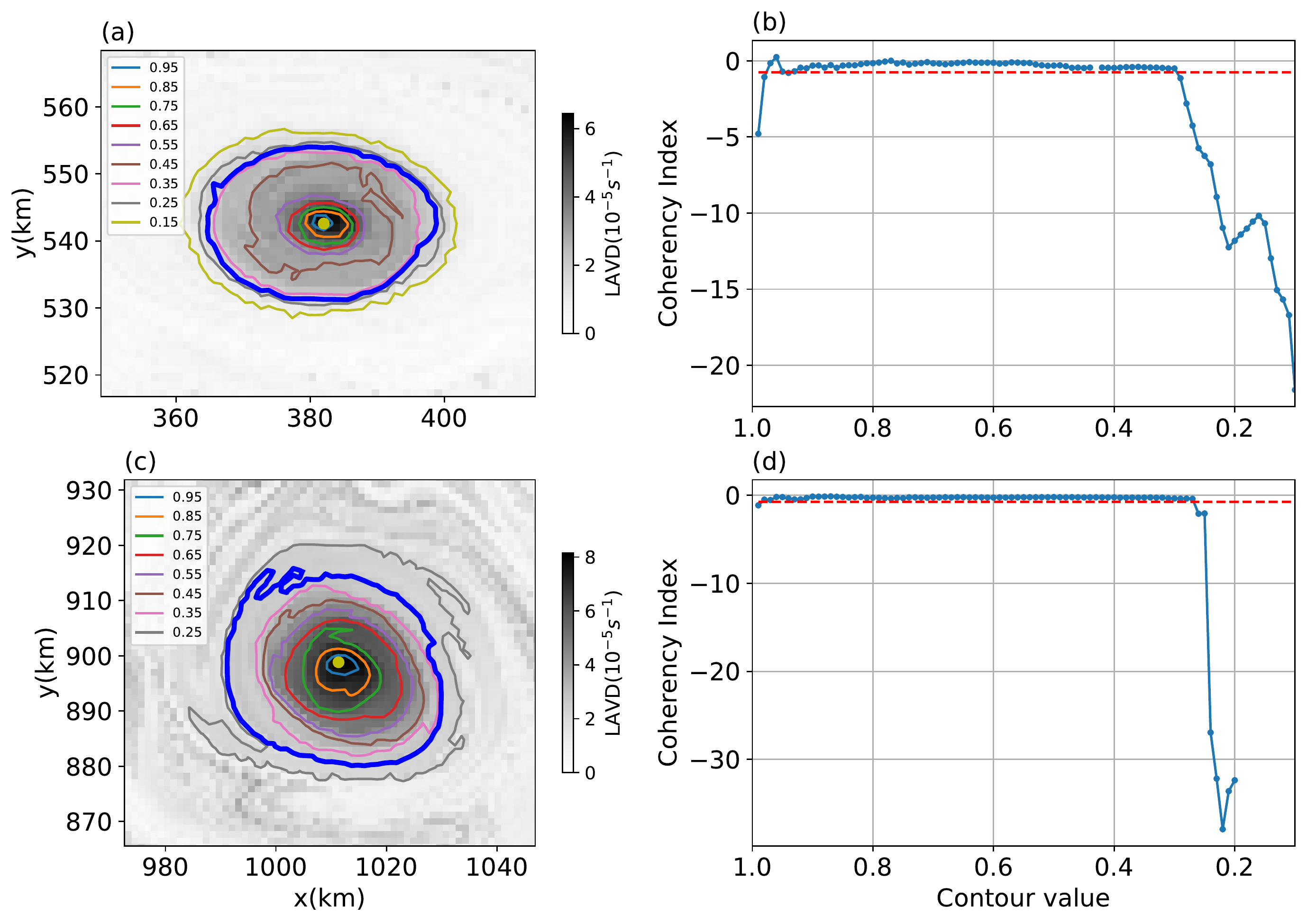}
\caption{(\textbf{a})  and (\textbf{c}) : LAVD fields at the initial particle positions for two example eddies (shading). Contours give values relative to the LAVD maximum at the eddy center with the outer boundary obtained from the threshold CI $>-0.75$ indicated by a thick blue contour. (\textbf{b})  and (\textbf{d}) : Radial distribution of the CI for the eddies shown in (\textbf{a})  and (\textbf{c}) , respectively, as a function of contour.  Dashed red lines indicate the threshold value for the CI.}   
\label{radial_ci}
\end{figure}

Figure \ref{radial_ci}a and \ref{radial_ci}c show example sets of LAVD contours around a local maximum where the LAVD field is plotted at the initial positions of particles at time $t_0$.  The outmost LAVD contour is determined by a threshold of a Coherency Index (CI).  CI measures the variation of spatial compactness of particles initially inserted within an LAVD contour and is defined as
\begin{linenomath}
\begin{equation}
\label{CI}
\mathrm{CI} = 1 - \frac{2}{R^2}\max_{t_1 \le t \le t_2}\delta^2(t)
\end{equation}
\end{linenomath}
where $\delta^2(t)=\langle |X(t) - \langle X(t) \rangle |^2 \rangle$ is the variance of particle positions inside an LAVD contour at time $t$, $X(t)$ is the position of the particle at time $t$, $t\in(t_0,t_1)$, $\langle \rangle $ indicates the average over all the particles inside the contour, and $| |$ is the standard Euclidean distance.  The Coherency Index compares the maximum of $\delta^2 (t)$ to $R^2/2$, which is the variance of the same number of particles in a disk with radius of $R$.  Since a disk is the most compact form of a particle cloud, any stretching or filamenting of the particle cloud will make CI negative.  The definition of CI in this study is different from the one defined by \citet{Tarshish_Abernathey_2018}, which measures the change of variance of particle positions at the final time relative to their initial variance.  This metric tends to favor large, spatially extended initial particle clouds that subsequently collapse into compact coherent eddies.  In this study, the normalization factor $R^2/2$ doesn\rq t depend on time or initial conditions and the particle spread is checked at each time step.  A detailed description of the advantages of this CI will be discussed in a forthcoming paper.  

Here we choose the threshold for CI as $-0.75$, which means we allow a maximum deviation from a disk of 75\%.  Figure \ref{radial_ci}b and \ref{radial_ci}d show the variation of CI on a nested set of LAVD contours for two example eddies.  The value of CI typically drops rapidly below the threshold value at certain contour level, which indicates that the enclosed eddy is no longer fully coherent.  Similar patterns are found in many examples, which suggests that this value is an approximate critical value below which the contour encloses incoherent particles.

In summary, the algorithm for detecting the coherent eddies within the LAVD field is the following: 
\begin{enumerate}[leftmargin=*,labelsep=4.9mm]
      \item Identify all the LAVD maxima with a minimum separation of 20 pixels (minimum separation suggested by \citet{Tarshish_Abernathey_2018}).
      \item Search from each maximum in LAVD for the outermost LAVD contour satisfying CI$>-0.75$ using the bisection method with an initial contour level of 0.36 times the LAVD value of the maximum.
      \item Remove eddies containing fewer than 200 particles (approximately equivalent to an area of 274 $\mathrm{km}^2$).  
\end{enumerate}
This algorithm is implemented by a Python package, \texttt{floater} modified from \citet{Abernathey_2018}.

\section{Eddy statistics}
\label{eddy_statistics_sec}
\subsection{Occurrence frequency of coherent eddies}
\unskip

\begin{figure}[H]
\includegraphics[width=\textwidth]{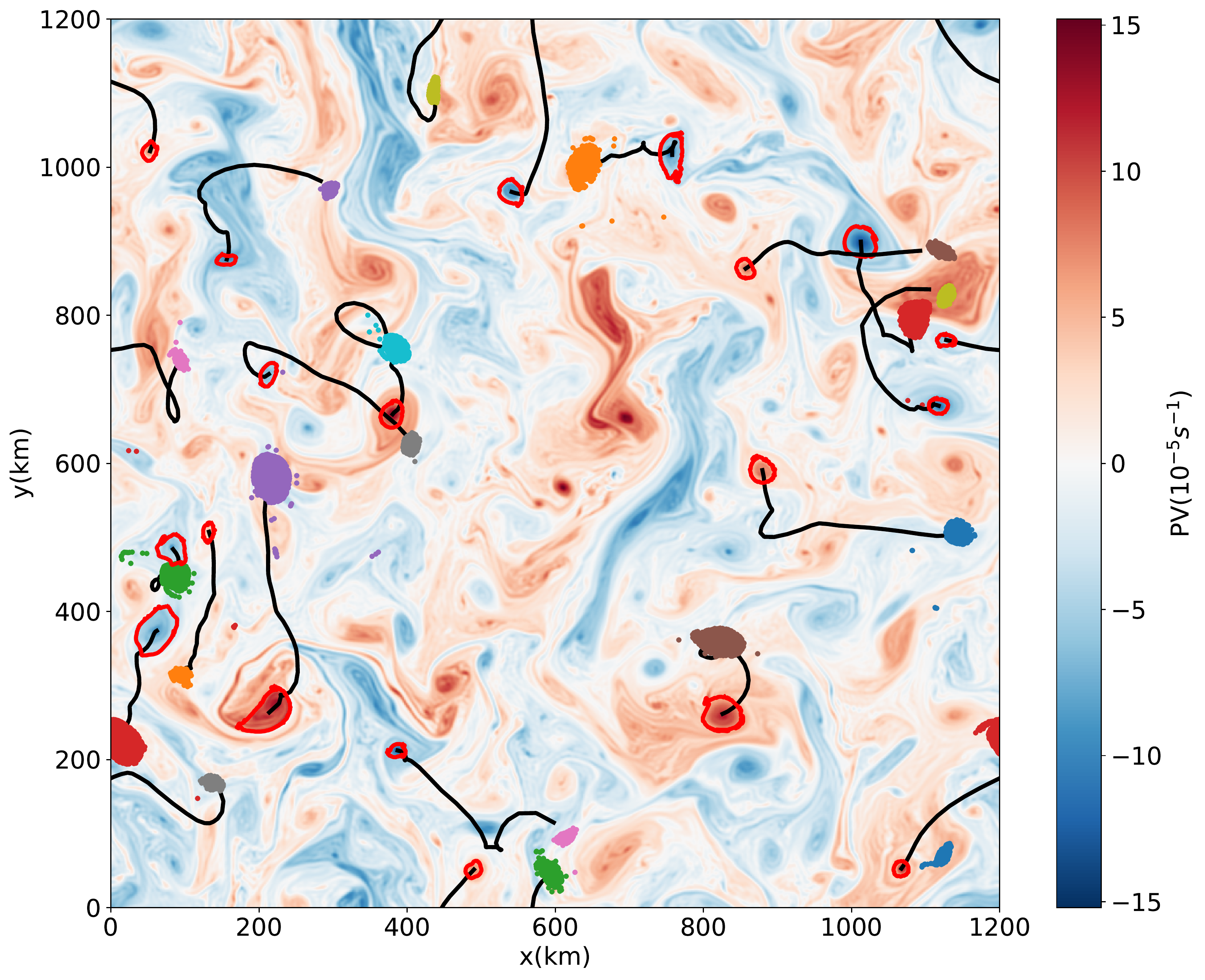}
\caption{An example set of 30-day eddies from the weak friction case.  The background field is the PV anomaly at the initial time, red curves give the boundaries of the coherent eddies at the initial time, black lines are the trajectories of eddy centers, and colored dots are the positions after 30 days of particles initially inside the eddies.}  
\label{snapshot_eddies}
\end{figure}

The method outlined in the previous section was used to detect 30-day, 60-day and 90-day coherent eddies in the first 30-, 60- and 90-day periods of each of the 100 90-day intervals.  The shortest interval are at least 5 times the eddy turnover time, estimated by $2\pi/\sqrt{Z}$, where $Z$ is the spatial mean eddy enstrophy in the upper layer.  The turnover times for strong, control and weak friction cases are 5.4, 4.6 and 3.5 days, respectively.  An example set of 30-day coherent eddies is shown in \figref{snapshot_eddies}, the particles inserted initially within the outer boundary of the eddies remain together in a compact shape after 30 days, with a few exceptions where eddies lose a small number of particles by the end of the 30-day period.

\begin{figure}[H]
\centering
\includegraphics[width=0.8 \textwidth]{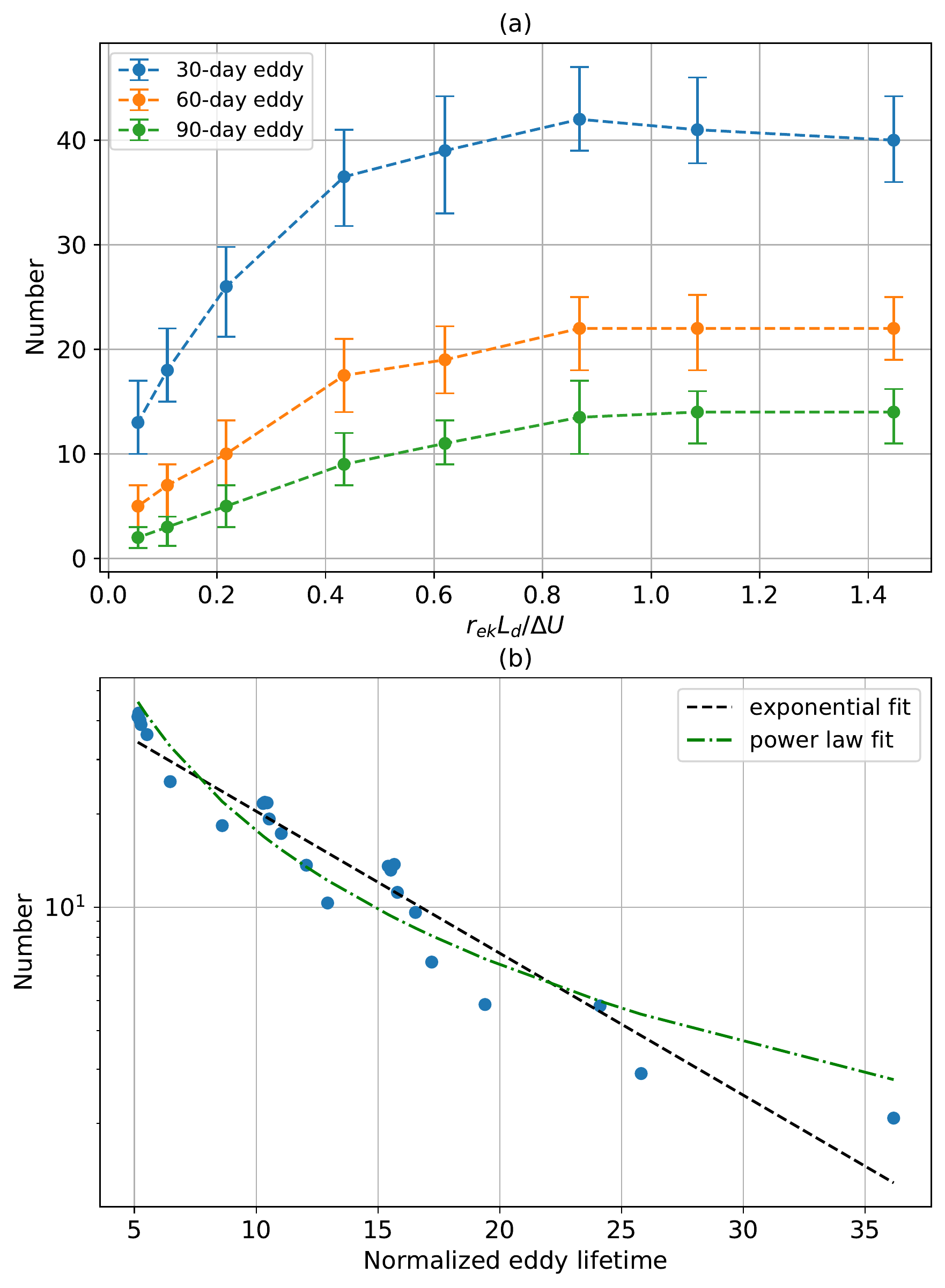}
\caption{(\textbf{a}) Number of coherent eddies in each time interval as a function of the nondimensional friction parameter $r_{ek} L_d/\Delta U$.  Blue, orange and green curves correspond to the median of the number of 30-, 60- and 90-day eddies, respectively.  Error bars span the 20th--80th percentiles over 100 time intervals.  (\textbf{b})  Average number of coherent eddies with lifetimes normalized by the eddy turnover time in each friction case.  The black dashed line is an exponential fit $N = 58.7 e^{-0.1 \tau}$ and green dash-dot line is a power law fit $N = 486 \tau^{-1.4}$, where $N$ is the number of coherent eddies and $\tau$ is the normalized lifetime.}  
\label{number_eddies}
\end{figure}

The number of coherent eddies in the 30-, 60- and 90-day intervals are shown in \figref{number_eddies}a as a function of friction parameter $r_{ek}$ nondimensionalized by intrinsic advection timescale $L_d/\Delta U$.  The coherent eddies are less numerous for smaller friction, while the coherent eddy frequency appears to plateau at large values of the friction parameter.  A possible interpretation is that the flow is more energetic for weak friction, so eddy interactions are more frequent and involve stronger strain fields; both of these effects would tend to make the eddies lose coherence more quickly.  Shorter-lived eddies are systematically more prevalent than longer-lived eddies.  Note that each 30-day period is included in the corresponding 60- and 90-day periods, so the 60- and 90-day eddies are subsets of the 30-day eddies.  In \figref{number_eddies}b the eddy lifetimes are normalized by the eddy turnover time in each friction case, and the number of coherent eddies versus their normalized lifetimes is given. Both an exponential fit and a power law fit for this relationship  have a p-value much smaller than 0.05, but the former one has a slightly smaller residual so is adopted by this study.  The number of coherent eddies decreases exponentially with an exponent of about $-0.1$ as the normalized lifetime increases, implying that an eddy has a nearly 10\% probability of losing coherence over each turnover time.  The decay rate of coherent eddies seems to be independent on friction and may be related to the nature of the turbulent flow itself.  Since the chance of an eddy loosing coherence is the same for every time interval, this implies that coherent eddies do not 'age'; that is, a recently-formed coherent eddy is as likely to loose coherence as a mature eddy.

\subsection{Eddy radius distribution}
\unskip

\begin{figure}[H]
\includegraphics[width=\textwidth]{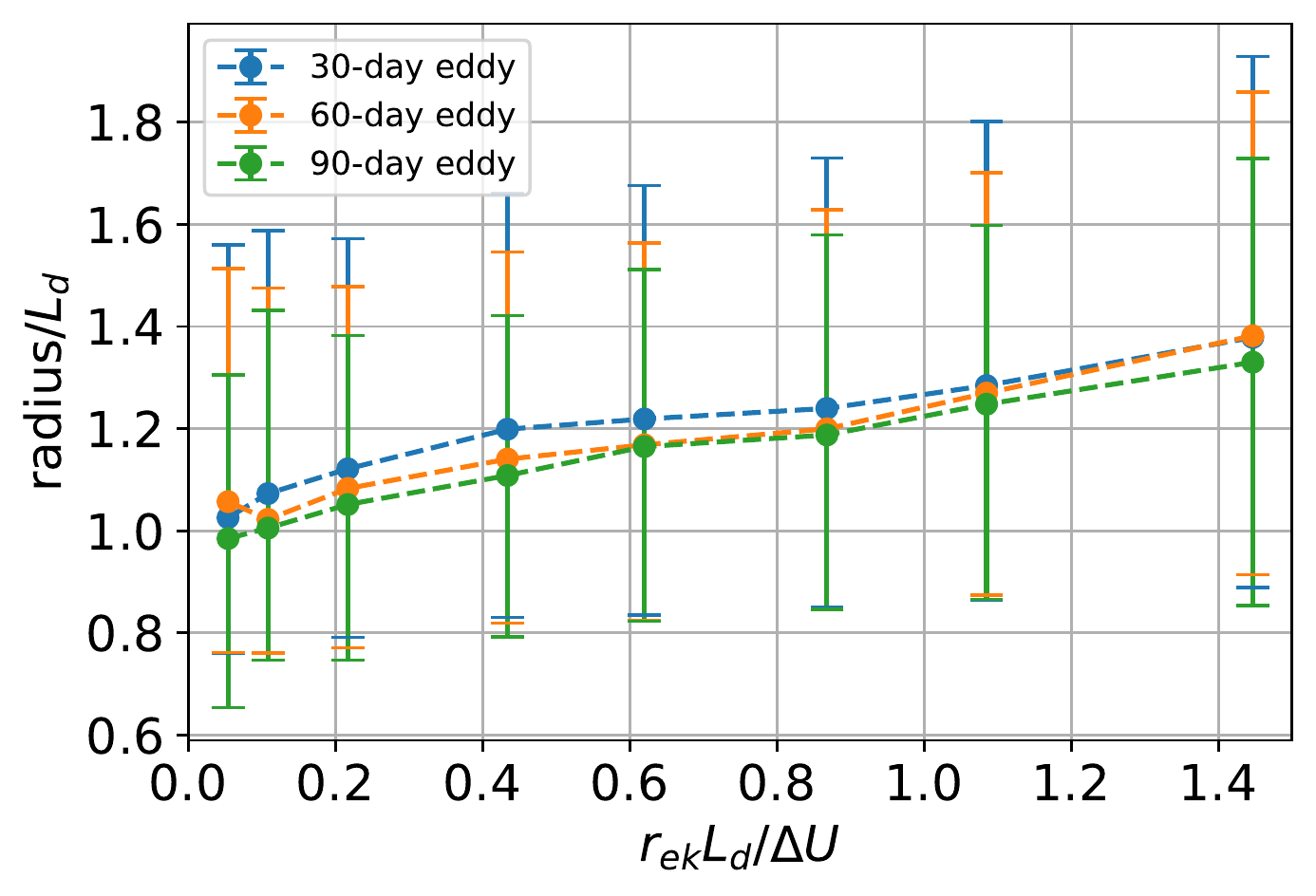}
\caption{Distribution of radius of coherent eddies as a function of the nondimensional friction parameter $r_{ek} L_d/\Delta U$.  Blue, orange and green dashed curves correspond to the median of the radius for 30-, 60- and 90-day eddies, respectively.  Error bars span the 20th--80th percentiles of the distribution.}   
\label{radius_distr}
\end{figure}

Eddy radius is estimated by $r=\sqrt{A/\pi}$, where $A$ is the area enclosed by the outer boundary of the eddy (i.e., the red curves in \figref{snapshot_eddies}).  The distribution of the eddy radii is given in \figref{radius_distr}.  The coherent eddies are generally small, with median radii slightly larger than the Rossby deformation radius $L_d$ (15 km).  Note that the radii shown here is the radii of the coherent core of the eddy, which is smaller than the eddy itself.  The median radius becomes slightly larger as the friction increases.  An alternative eddy radial scale is the e-folding scale of its PV, which is estimated to be around 1.2 times the radius of the coherent core identified by the LAVD method.

\subsection{Radial structure}
\unskip

The radial structure of the vorticity of isolated vortices in a two-dimensional viscous flow has been shown to have an analytical form
\begin{linenomath}
\begin{equation}
\label{radial_function}
\omega_n=\left(1 - \frac{r_n^2}{2}\right) \exp{\left(-\frac{r_n^2}{2}\right)},
\end{equation} 
\end{linenomath}
where $\omega_n=\omega/\omega_0$ and $r_n=r/r_0$ are vorticity anomaly and radial distance normalized by the amplitude $\omega_0$ at eddy center and radius $r_0$ of the corresponding eddies.  The form given in \eqref{radial_function} is a similarity solution for the viscous decay of initially compact eddies with zero net vorticity; it was derived by \citet{Taylor_1918}, observed in laboratory experiments \cite{Trieling_Heijst_1998}, and has been adopted to describe the radial structure of pressure anomaly---instead of vorticity---for ocean mesoscale eddies observed in satellite and Argo observations \cite{Zhang_Zhang_2013}.

\begin{figure}[H]
\centering
\includegraphics[width=0.6 \textwidth]{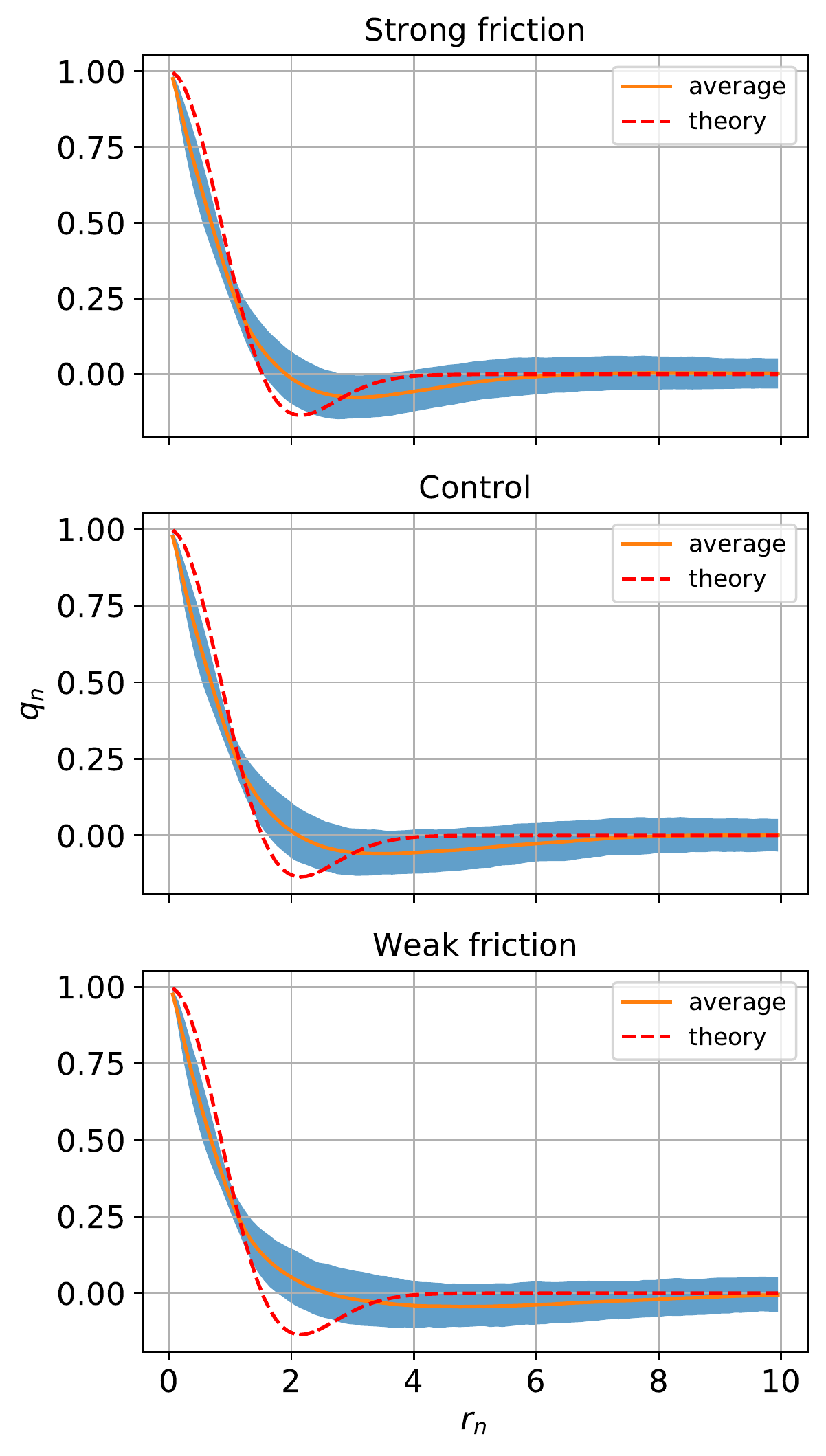}
\caption{Mean radial structure of PV in coherent eddies. Panels from top to bottom are from the strong, control and weak friction cases. The orange solid curve is the PV normalized by its maximum value in the eddy center and averaged over all the coherent eddies. The blue shading spans the 20th--80th percentiles of the normalized PV at each radial distance normalized by the e-folding scale of the PV distribution. The red dashed line is the theoretical radial structure given by \eqref{radial_function} and is normalized by its own e-folding scale. }  
\label{radial_PV}
\end{figure}

We compare the radial structure of the coherent eddies to \eqref{radial_function}, using PV instead of vorticity because the PV is the QG analog of vorticity in two-dimensional flow.  The PV and radial distance $r$ are normalized by $q_0$ and $r_0$, respectively, where $q_0$, is the PV anomaly at the center of the eddy, and the scale parameter $r_0$ is the e-folding scale of the PV distribution of the eddy.  The radial distribution of each eddy is obtained by azimuthally averaging in contiguous radial windows of length of 0.1 $r_0$.  The average normalized radial PV structure for all the eddies is shown in \figref{radial_PV}.  The radial PV structure has a negative lobe similar to the theoretical function \eqref{radial_function} but is steeper near the eddy center, and the minimum is shifted away from the origin.  The negative lobe indicating the opposite-sign PV rings around the eddies becomes smaller as the friction increases.

\subsection{Eddy propagation}
\unskip
\subsubsection{Eddy zonal propagation velocity}
\label{sec_zonal_speed}

\begin{figure}[H]
\includegraphics[width=\textwidth]{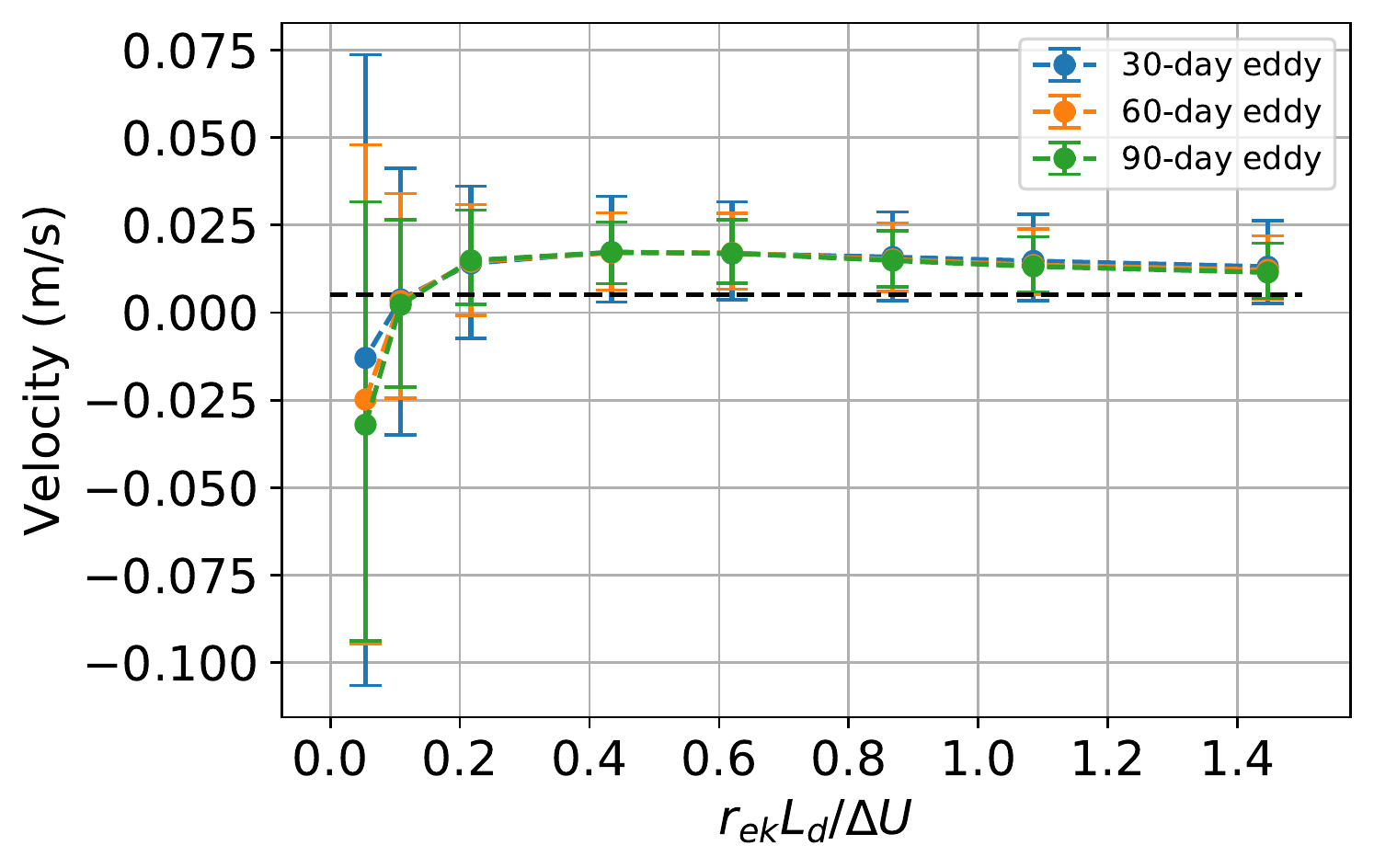}
\caption{Distribution of zonal propagation velocity of coherent eddies as a function of the nondimensional friction parameter $r_{ek} L_d/\Delta U$. Blue, orange and green curves correspond to the median for 30-, 60- and 90-day coherent eddies, respectively.  Error bars span the 20th--80th percentiles of the distribution.  The dashed black line indicates the baroclinic Rossby wave speed $c_{BC}$.} 
\label{zonal_speed}
\end{figure}

The mean zonal propagation velocity of coherent eddies in three friction cases is estimated by their zonal displacements over the corresponding time interval.  Studies have shown that the eddy zonal phase speeds estimated from satellite observations are predicted well by the linear baroclinic Rossby wave theory \cite{McWilliams_Flierl_1979, Klocker_Abernathey_2014}.  In the long-wave limit and taking account of the Doppler shift by the mean flow \cite{Klocker_Abernathey_2014, Wang_Jansen_2016}, the baroclinic Rossby wave speed is estimated as 
\begin{linenomath}
\begin{equation}
\label{c_bc}
c_{BC}=U_b - \beta L_d^2 = 0.51 \mathrm{cm/s} 
\end{equation}
\end{linenomath}
where $U_b = (\delta U_1 + U_2) / (1 + \delta)$ is the depth-averaged mean flow.  Figure \ref{zonal_speed} shows the zonal propagation velocities of coherent eddies are within one standard deviation of $c_{BC}$, although the velocities are systematically larger than $c_{BC}$ for all but the two smallest values of the friction parameter.  This may be due to the tendency of the eddies to concentrate in the faster upper layer when friction is strong \cite{Wang_Jansen_2016}.

\subsubsection{Eddy meridional displacement}
\unskip
\label{sec_meri_disp}

\begin{figure}[H]
\centering
\includegraphics[width=0.9 \textwidth]{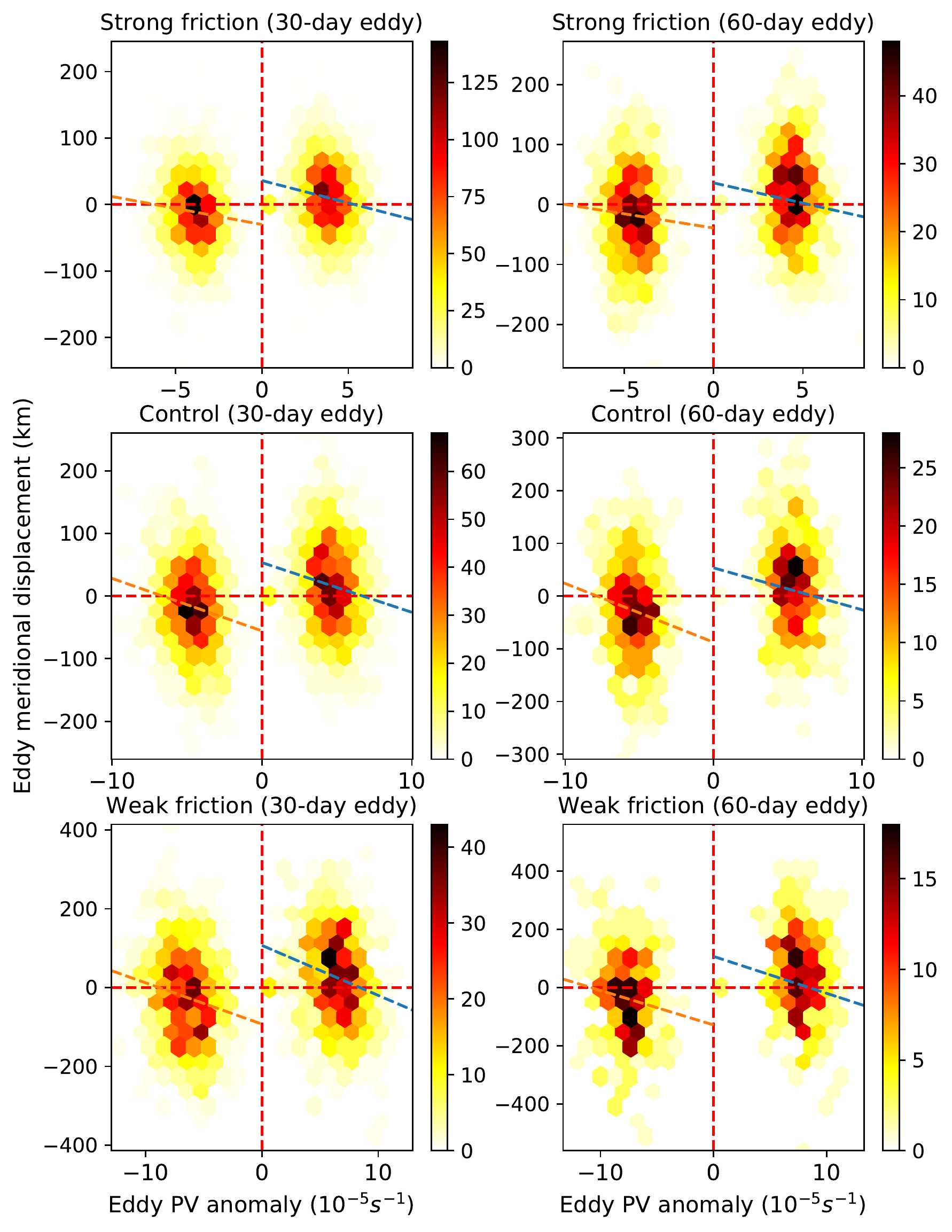}
\caption{Two-dimensional histogram of the joint distribution of the eddy PV anomaly and meridional displacement for (left) 30-day and (right) 60-day coherent eddies in the (top) strong, (middle) control, and (bottom) weak friction cases. The shading is proportional to the number of coherent eddies in each bin. Red and blue dashed lines indicate the linear regression lines for the negative and positive cluster of points, respectively.} 
\label{meri_speed}
\end{figure}

The meridional displacement ($Y$) of the 30- and 60-day coherent eddies versus their temporal and spatial mean PV anomalies are shown in \figref{meri_speed}.  The left and right cluster of high-density bins represent the anticyclones and cyclones.  The meridional displacement is correlated with the sign of the PV anomaly, with cyclones (anticyclones) tending to drift northward (southward).  A two-sample t test shows that the means of meridional displacement of cyclones and anticyclones are significantly different ($p\ll0.05$).  In all three cases approximately 60\% of 30-day cyclones propagate poleward and 60\% of 30-day anticyclones propagate equatorward.  The identical but reversed percentages in all the three cases illustrates the symmetry between cyclones and anticyclones in QG system.  The magnitude of meridional displacement of coherent eddies has a negative linear relationship with their magnitude of PV, implying that stronger coherent eddies propagate more slowly in meridional direction.  The same relationship was also observed   in a barotropic QG model \cite{Early_Samelson_2011}, but there is, as yet, no straightforward explanation for this relationship.

The preference for cyclones and anticyclones to drift in opposite directions has been observed in a global investigation of ocean mesoscale eddies \cite{Chelton_2011}, which indicates that this is a universal phenomenon for ocean eddies.  Indeed, \citet{McWilliams_Flierl_1979} demonstrated that the secondary circulation induced by the leading and trailing streamfunction anomalies of an anticyclone acts to advect it southward.  They showed that nonlinearity is essential for this self-advection process, and beta effect acts to induce the initial streamfunction dispersion.  Similar mechanism is also discussed by \citet{Cushman_2011} with clear schematics.  A more fundamental interpretation is that a vortex has tendency to return to a rest latitude where the ambient PV matches that of the eddy \cite{Rossby_1949, McWilliams_Flierl_1979}.  Since this meridional drift reduces the eddy's PV anomaly, the drift leads to decay of the eddy and may lead to loss of coherence.  Whether this effect is a significant decay mechanism still requires quantitative assessment. 

\section{PV transport by coherent eddies}
\label{sec_PV_trans}
\subsection{Transport by trapping}
\unskip
The Lagrangian trajectories of the coherent eddies can be used to estimate their PV transport.  The total Lagrangian advective PV flux in the meridional direction is
\begin{linenomath}
\begin{equation}
\label{Q_PV}
Q_t = \overline{v'q'},
\end{equation}
\end{linenomath}
where $v'$ and $q'$ are the velocity and PV anomalies for each Lagrangian particle and the overbar is an ensemble average over all the Lagrangian particles during each 30-, 60- or 90-day period.

\begin{figure}[H]
\includegraphics[width=\textwidth]{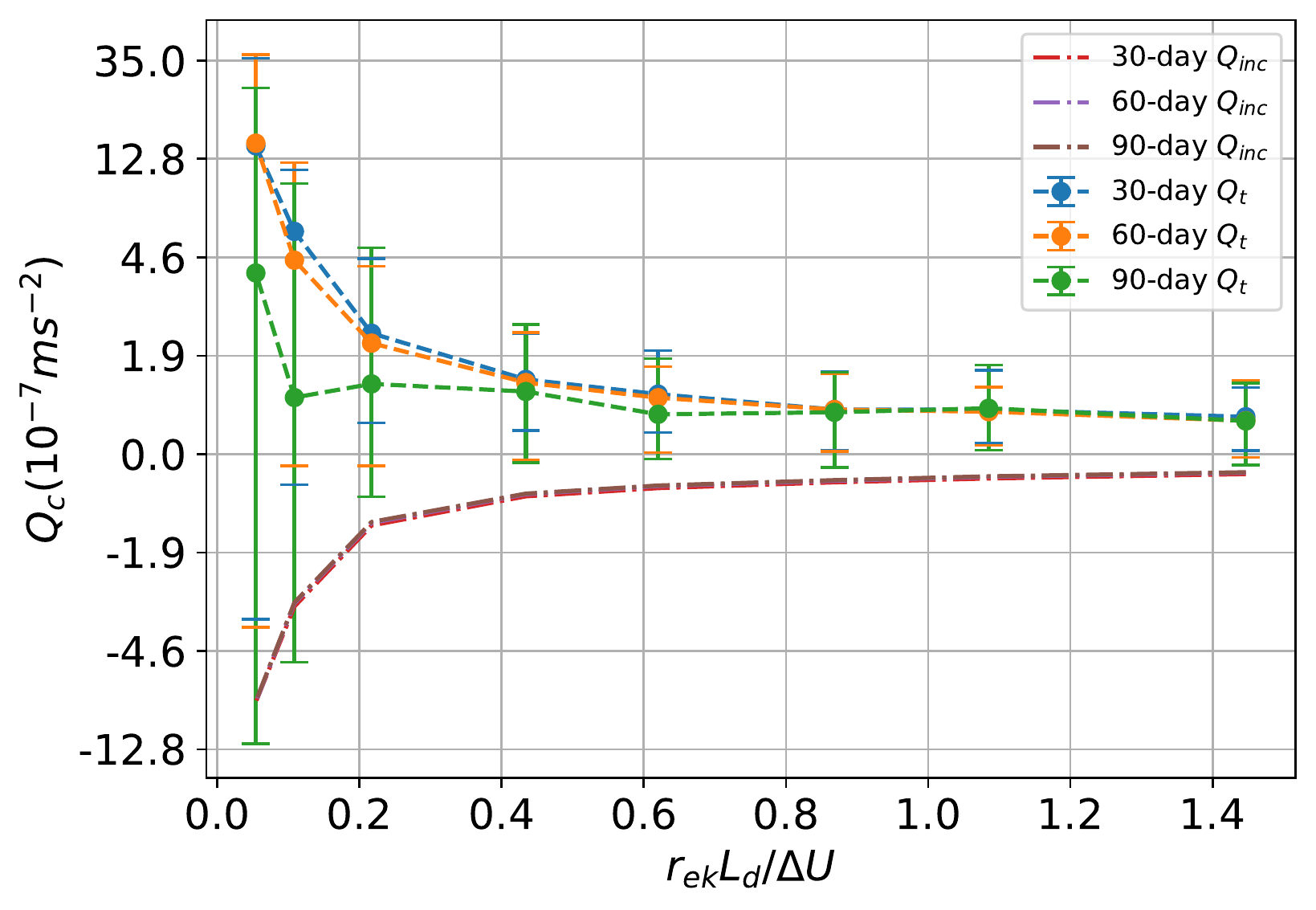}
\caption{Net meridional PV flux (per coherent particle) due to coherent eddies as a function of the nondimensional friction parameter $r_{ek} L_d/\Delta U$.  Blue, orange and green dashed curves correspond to the median of PV flux for 30-, 60- and 90-day coherent eddies, respectively.  Error bars span the 20th--80th percentiles over 100 time intervals.  The dash-dot red, purple and grey lines indicate the averaged incoherent meridional PV flux over the 100 30-, 60- and 90-day intervals, respectively.}  
\label{coh_flux}
\end{figure}

The PV flux due to the coherent eddies is
\begin{linenomath}
\begin{equation}
\label{Q_c}
Q_c = \frac{\overline{Av'q'} }{\overline{A}},
\end{equation}
\end{linenomath}
where $A$ is a masking function that is 1 for particles inside coherent eddies and 0 outside. $\overline{A}$ is
the fractional area occupied by coherent eddies, so $Q_c$ is the mean flux per coherent particle due to coherent eddies---the total PV flux due to coherent eddies is obtained by multiplying \eqref{Q_c} by $\bar{A}$.  The PV flux per particle due to incoherent motions is
\begin{linenomath}
\begin{equation}
\label{Q_inc}
Q_{inc} = \frac{\overline{(1-A)v'q' }}{\overline{1-A}} 
\end{equation}
\end{linenomath}
where $\overline{1-A}$ is fraction of the domain outside of coherent eddies. Figure \ref{coh_flux} shows the statistics of coherent PV flux, $Q_c$, calculated in each of the 30-, 60- and 90-day periods and the corresponding time averaged incoherent PV flux, $Q_{inc}$, which is close to the total PV flux since the area occupied by coherent eddies is small.  The coherent PV flux $Q_c$ is systematically positive (upgradient), while the mean incoherent PV flux $Q_{inc}$ is always negative (downgradient).  This means that the PV transport inside coherent eddies is systematically opposite to the transport in the background flow, implicating that coherent eddies play a special dynamical role.  

The total meridional PV flux by coherent eddies is $\overline{A}Q_c$, so the fraction of the total flux due to coherent eddies is $\overline{A}Q_c/Q_t$.  Figure \ref{PV_trans}a shows the median of the ratio $\overline{A}Q_c/Q_t$ is systematically less than 10\% (upper limits are less than 20 \%) even in very high friction cases.  Thus, the PV transport by coherent eddies is small compared to the total PV transport, but is systematically in the opposite direction.

The upgradient coherent PV flux is due to the systematically poleward translation of cyclones and equatorward translation of anticyclones (cf. section \ref{eddy_statistics_sec}).  Since the total PV transport must be downgradient in statistical equilibrium, the downgradient incoherent PV transport over the whole domain must be much larger to compensate for the upgradient transport due to coherent eddies.  This sets a dynamical constraint on the contribution of PV transport due to coherent eddies.  However, were flows less zonally uniform, advection of enstrophy by mean flow might become important \cite{Marshall_1984}.  \citet{Holland_Rhines_1980} showed that strong advection of enstrophy by gyres leads to both up- and down-gradient PV fluxes.  Thus, the upgradient PV transport by coherent eddies might be more significant in regions with strong nonuniform mean flows.

\subsection{Transport by stirring}
\unskip

The small contribution of PV transport inside coherent eddies does not necessarily mean that coherent eddies are not important in meridional PV transport.  In addition to trapping fluid inside their cores, coherent eddies also induce the flow in the far field.  Transport by these flows can also be thought of as ``due to'' the coherent eddies.  Studies of two-dimensional turbulence have proposed a two-component fluid view \cite{Bracco_LaCasce_2000, Provenzale_Babiano_2008} for modeling flow with coherent vortices immersed in background turbulence.  The main idea for this view is to decompose the Eulerian flow into two dynamical components: one is generated by coherent vortices and the other is generated by the background vorticity field \cite{Pasquero_Provenzale_2001}.  This allows us to estimate the transport by the flow induced by coherent eddies including the far field.

\begin{figure}[H]
\includegraphics[width=\textwidth]{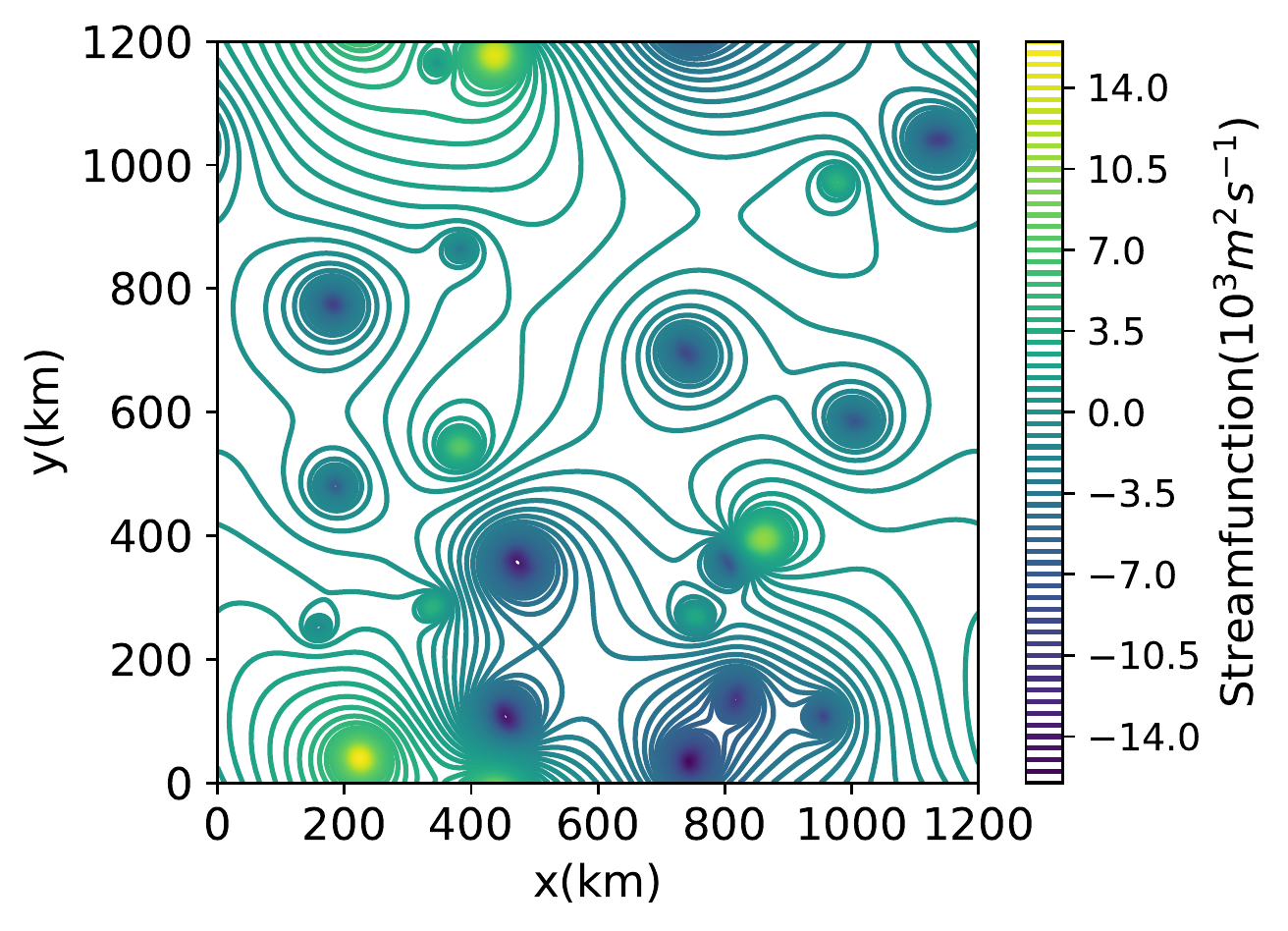}
\caption{An example of flow field $\psi_c$ induced by coherent eddies. } 
\label{ppvi_strf}
\end{figure}

The flow induced by the coherent eddies is calculated using piecewise PV inversion (PPVI) \cite{Hoskins_1985, Davis_1992, Egger_2008}.  PPVI has been widely used in diagnosing the effect of portions of the PV in geophysical systems, such as effect of a PV anomaly at the  tropopause on the cyclogenesis \cite{Eliassen_Kleinschmidt_1957}.  The PV inversion process is linear in QG, so the total flow is the sum of flows induced by all the portions of the PV.  Specifically, the streamfunction for the total flow is decomposed as
\begin{linenomath}
\begin{equation}
\label{ppvi_psi}
\psi = \psi_c + \psi_{inc}
\end{equation}
\end{linenomath}
where $\psi_c$ and $\psi_{inc}$ are the flow generated by the coherent eddies and the background PV field, respectively.  The former $\psi_c$ is calculated by the inversion of the PV field $q_c$ which is nonzero only in the upper layer inside coherent eddies.  The incoherent flow, $\psi_{inc}$, is derived from the remaining PV field, including the lower layer.  An example of the flow field $\psi_c$ is shown in \figref{ppvi_strf}---the PV inside coherent eddies induce the flow throughout the whole domain.  

\begin{figure}[H]
\centering
\includegraphics[width=0.8 \textwidth]{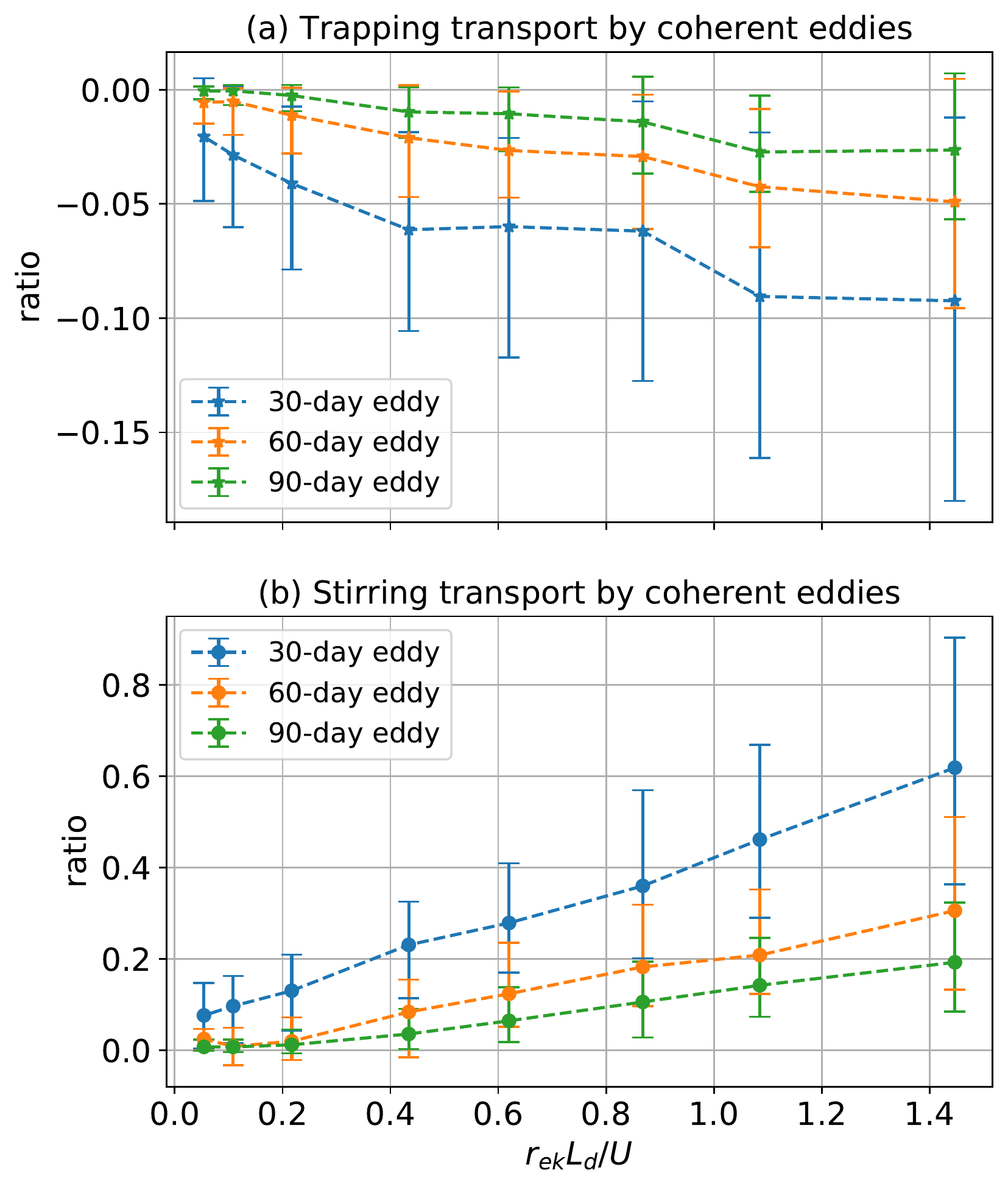}
\caption{(\textbf{a}) Ratio of meridional PV transport due to trapping of coherent eddies to total PV transport as a function of the nondimensional friction parameter $r_{ek} L_d/\Delta U$.  (\textbf{b}) Ratio of meridional PV transport by the flow induced by coherent eddies to total PV transport as a function of the nondimensional friction parameter $r_{ek} L_d/\Delta U$.  In both panels, blue, orange and green curves correspond to the median for 30-, 60- and 90-day coherent eddies, respectively, and the error bars span the 20th--80th percentiles over 100 time intervals. }  
\label{PV_trans}
\end{figure}

The meridional PV flux due to $\psi_c$ is $\overline{v_cq}$, where $q$ is the total (coherent plus incoherent) PV field.  The PV transport by the flow induced by coherent eddies is 10-60\% of the total PV transport, with stronger friction associated with higher ratios (\figref{PV_trans}b).  In addition, this transport is systematically downgradient so the net PV transport due to both the stirring and trapping effect of coherent eddies is downgradient.  The meridional PV transport due to coherent eddies is therefore primarily due to stirring rather than the drift of their cores, but these two components are actually correlated with each other.  More than 80\% of the induced PV transport is localized within 3 radii of the coherent cores; in this region, more than $50$\% of the EKE is induced by the PV inside the eddy cores.

\section{Tracer transport by coherent eddies}
\label{diff_sec}

The PV distribution is highly correlated with the eddies---indeed, the eddies are PV extrema almost by definition.  The role of coherent eddies in the transport of a ``generic'' passive tracer (i.e., a tracer whose initial distribution is independent of the distribution of the eddies) is also of interest.

\begin{figure}[H]
\centering
\includegraphics[width=0.7 \textwidth]{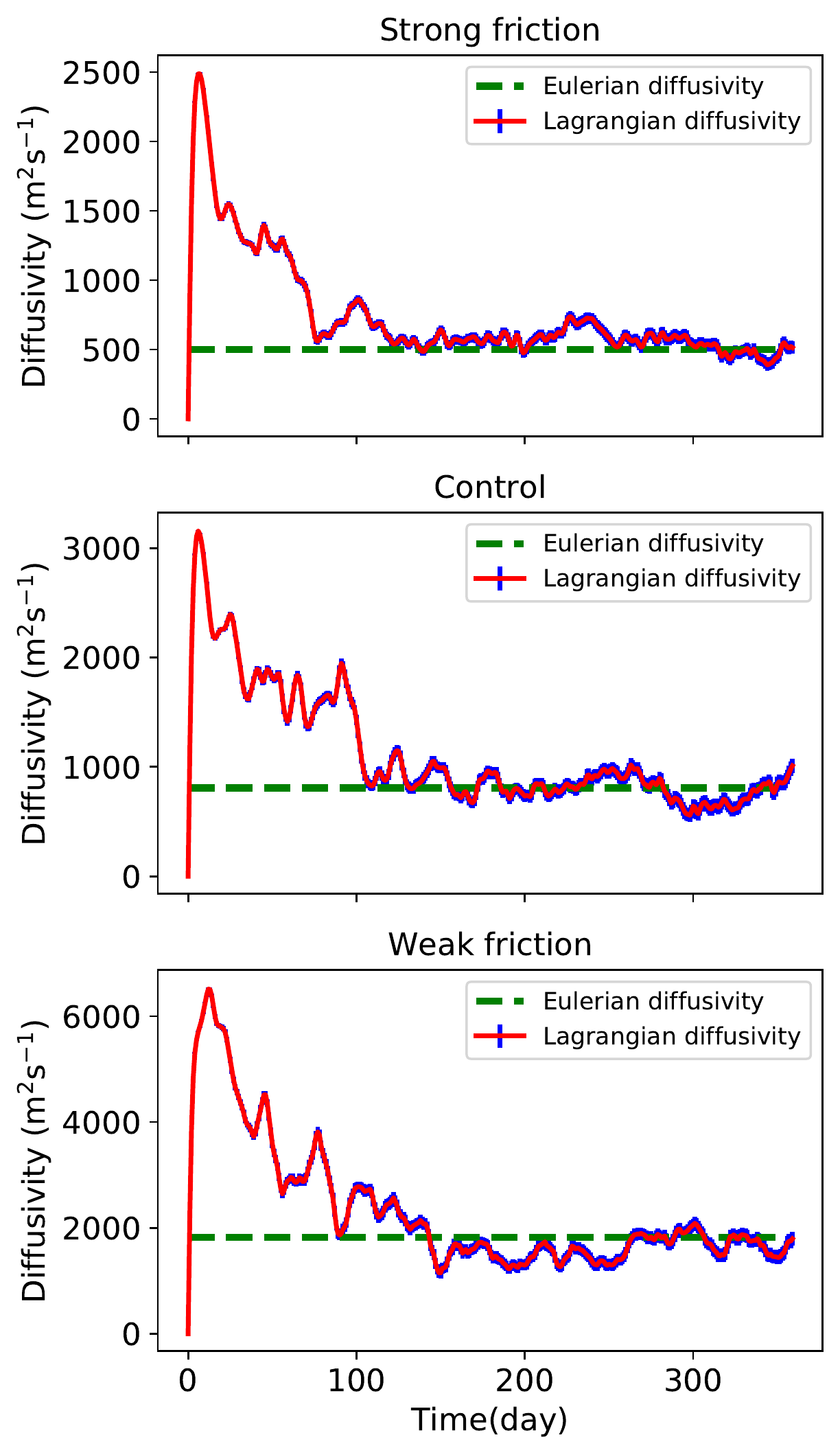}
\caption{Upper level Lagrangian diffusivity calculated by \eqref{K_Lagr} (red curve) with shaded error bars of 2 times standard error, and Eulerian diffusivity (green dashed curve).}  
\label{Lagr_diff}
\end{figure}

The single-particle Lagrangian diffusivity is calculated by \eqref{K_Lagr}  in a 360-day period.  The time series of the Lagrangian diffusivity in the three friction cases are shown in \figref{Lagr_diff} and compared to the Eulerian meridional PV diffusivity
\begin{linenomath}
\begin{equation}
\label{Eulerian_diffusivity}
K = -\frac{\overline{v'q'}}{\overline{q_y}}
\end{equation}
\end{linenomath}
where an overbar indicates an area and time average.  The Eulerian diffusivity becomes larger as the friction weakens.  The Lagrangian diffusivity of all the three simulations increases to a maximum within the first 10 days, and then decays and asymptotes to a constant value which is close to the domain-mean Eulerian diffusivity.

We can use a small modification of the Taylor diffusivity \eqref{K_Lagr} to calculate the coherent diffusivity; that is, the diffusivity due to coherent eddies.  The modified diffusivity is
\begin{linenomath}
\begin{equation}
\label{K_coh}
K_C (y_0,t) = \frac{1}{2}  \frac{d}{dt} \left[\frac{1}{N\overline{A}} \sum_{i=0}^{N-1}A_i (y_i(t) - y_{i0} )^2\right],
\end{equation}
\end{linenomath}
where $A_i$ is a masking function which is unity for particles inside coherent eddies and zero otherwise.

\begin{figure}[H]
\centering
\includegraphics[width=0.7 \textwidth]{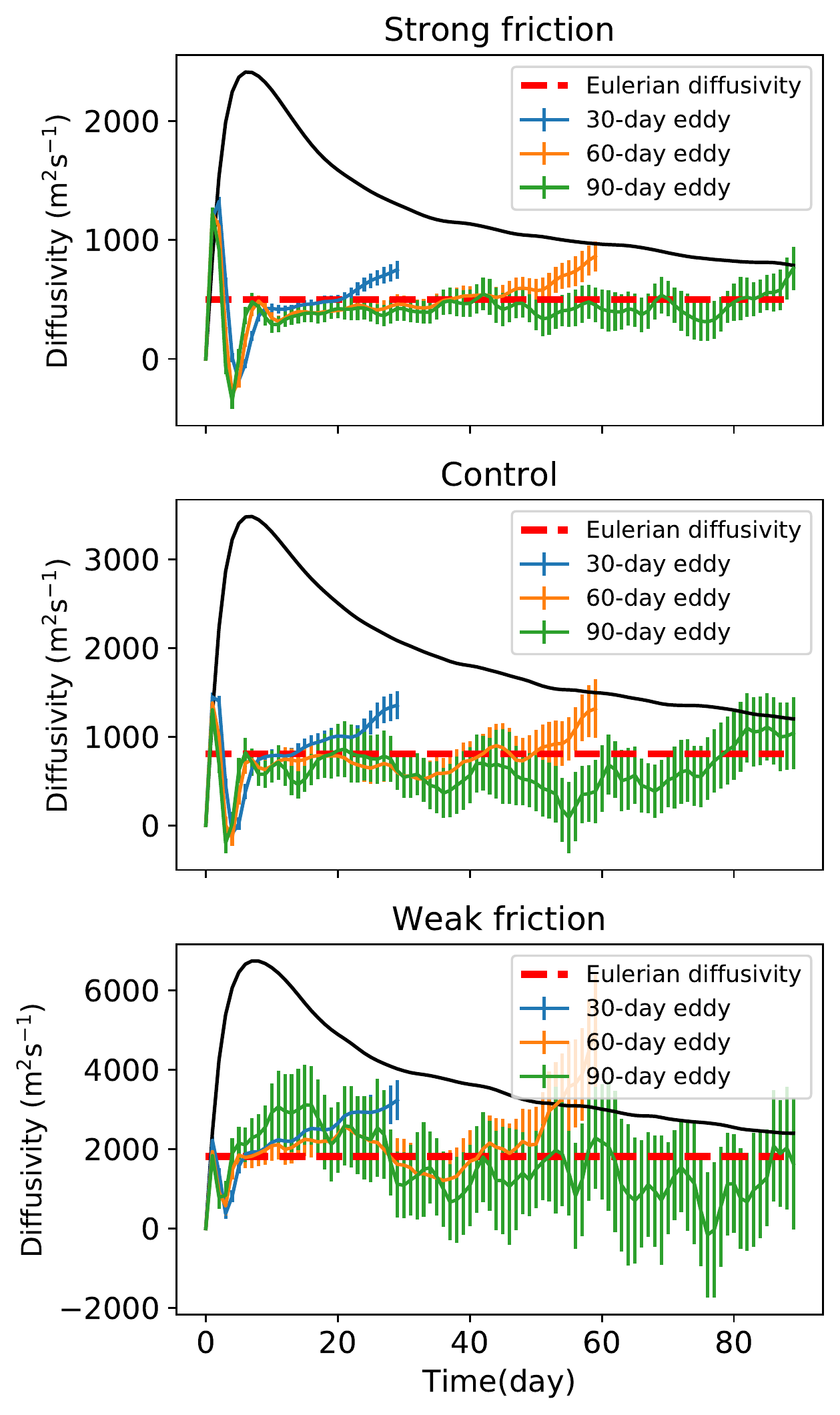}
\caption{Time series of coherent diffusivity (per coherent particle) in three simulations. Blue, orange and green lines are 30-, 60- and 90-day coherent diffusivities, respectively.  Error bars give two times the standard error.  Black lines indicate the Lagrangian diffusivity averaged over all the particles in the same time period. Red dashed line indicates the Eulerian diffusivity.}  
\label{coh_diff}
\end{figure}

The coherent diffusivity increases to a maximum within the first 1--2 days, then drops to a local minimum near the 5th day, and then increases again (\figref{coh_diff}).  The period of this diffusivity oscillation is on the same order of the eddy turnover time, so may be due to the swirling motion of the eddies.  After this oscillation, the coherent diffusivity converges to the Eulerian diffusivity much faster than the Lagrangian diffusivity averaged over all the particles. However, this does not mean the coherent diffusivity represents the total Lagrangian diffusivity. The contribution of coherent eddies to the total diffusivity is $\overline{A}K_C/K$, which is the same as the `coherent diffusivity' defined in \citet{Abernathey_Haller_2018}. Indeed, the fraction of Lagrangian diffusivity due to coherent eddies is less than 2\% for the three friction cases shown in \figref{coh_diff}.  The estimate for 30-day eddies is slightly larger than the estimate by \citet{Abernathey_Haller_2018} but is still not a significant contribution to the total diffusivity.  The smallness of the coherent diffusivity is due to the small fractional area, $\overline{A}$, occupied by coherent eddies.

The coherent diffusivity is mostly positive, which is inconsistent with the upgradient coherent PV flux discussed in section \ref{sec_PV_trans}.  This implies that the PV transport by the coherent cores is not a diffusion process and isn't captured merely by the dispersion of Lagrangian particles in the cores. Taylor's \cite{Taylor_1922} derivation assumes that the initial tracer gradient is uniform, and that the particle motion is independent of the tracer distribution.  This is clearly not the case for PV transport by coherent eddies since the eddies themselves are the PV extrema and their motion depends on the sign of their PV.  Taylor's theory therefore is not applicable to PV transport by coherent eddies since its assumptions are violated.

\section{Discussion and conclusions}
\label{discussion_sec}

This study uses the objective Lagrangian coherent eddy detection method of \citet{Haller_Hadjighasem_2016} to identify the coherent eddies in two-layer quasigeostrophic turbulence.  Materially coherent eddies are detected for a wide range of parameters, with stronger friction associated with a greater prevalence of coherent eddies.  The coherent eddies take the form of a small (near the deformation radius) materially coherent core imbedded in a larger area of PV anomaly characterized by incoherent particle motion.

The cyclones and anticyclones have distinct preferences for meridional propagation.  Cyclones systematically propagate poleward and anticyclones propagate equatorward, which gives rise to upgradient PV transport.  This upgradient PV transport is the opposite of the general PV transport by the background flow and is not captured by the Lagrangian diffusivity of \citet{Taylor_1922}.  This implicates a special dynamical role played by the coherent eddies, which is distinct from the diffusion processes.  The PV transport due to the drift of coherent eddy cores is small compared to the total PV transport, but these coherent cores can induce the flow in the far field which gives rise to significant downgradient PV transport in the periphery of the eddy cores.  The former component of the transport is due to trapping, while the latter is due to stirring.  This study shows that the stirring transport by coherent eddies is more likely to be parameterized as a diffusion process, while the trapping transport---although itself small---might determine the evolution of eddies themselves.  How the trapping and stirring transport correlate with each other and how they are correlated with the formation and decay of coherent eddies needs further study.

The meridional propagation tendency of coherent eddies relies on the $\beta$-effect, and PV is itself an active tracer, which makes the PV transport special.  It would also be interesting to investigate the role of coherent eddies in the transport of tracers like heat and biochemical constituents which have been shown to correlate with the signs of PV of mesoscale eddies in the ocean.  \citet{Thompson_Young_2006} investigated the eddy heat flux in a two-mode, \textit{f}-plane QG model and proposed that the downgradient heat flux is due to the systematic propagation of hot anticyclones poleward and cold cyclones equatorward, which is opposite to our cases.  However, \citet{Dong_McWilliams_2014} showed that the heat transport by mesoscale eddies in the ocean is significantly upgradient near the tropics.  It would be worthwhile to investigate this process in the $\beta$-plane QG model.  

The flows considered in this study are statistically homogeneous, while the transport and mixing in ocean gyres are anisotropic and inhomogeneous \cite{Berloff_McWilliams_2002a}.  Previous studies on oceanic gyres have shown that tracer transport by eddies can differ from homogeneous case in many ways, with subdiffusive single-particle dispersion due to coherent structures \cite{Berloff_McWilliams_2002a}, mixing nonlocality within jets \cite{Chen_Waterman_2017} and the transport barriers in eastward jets.  Whether the trapping or stirring transport of coherent eddies is significant in such systems and how coherent eddies interact with the zonal jets or meridional western boundary currents would be interesting to explore.  Better understanding of the universal transport properties of coherent eddies will improve our ability to diagnose the turbulent flow structures and parameterize the eddy fluxes in coarse-resolution climate models.

\vspace{6pt} 



\authorcontributions{conceptualization, R.P.A., C.L.P.W. and W.Z.; methodology, R.P.A.; software, R.P.A., W.Z. and C.L.P.W.; validation, W.Z., C.L.P.W. and R.P.A.; formal analysis, W.Z. and C.L.P.W.; investigation, W.Z. and C.L.P.W.; resources, R.P.A. and C.L.P.W.; data curation, W.Z.; writing--original draft preparation, W.Z.; writing--review and editing, C.L.P.W. and R.P.A.; visualization, W.Z. and C.L.P.W.; supervision, C.L.P.W. and R.P.A.; project administration, C.L.P.W..}

\funding{This research received no external funding.}

\acknowledgments{We thank Stony Brook Research Computing and Cyber Infrastructure and the Institute for Advanced Computational Science at Stony Brook University for access to the high-performance SeaWulf computing system, which is supported by the NSF.  We also thank John Marshall for suggesting the piecewise PV inversion method.}

\conflictsofinterest{The authors declare no conflict of interest. The funders had no role in the design of the study; in the collection, analyses, or interpretation of data; in the writing of the manuscript, or in the decision to publish the results.} 

\abbreviations{The following abbreviations are used in this manuscript:\\

\noindent 
\begin{tabular}{@{}ll}
MDPI & Multidisciplinary Digital Publishing Institute\\
LAVD & Lagrangian-averaged vorticity deviation\\
PV & potential vorticity\\
RCLV & rotationally-coherent Lagrangian vortex\\
QG & quasigeostrophic\\
EKE & eddy kinetic energy\\
CI & coherency Index\\
PPVI & piecewise PV inversion\\
R.P.A & Ryan Abernathey\\
C.L.P.W & Christopher L.P. Wolfe\\
W.Z. & Wenda Zhang\\
NSF & National Science Foundation
\end{tabular}}

\appendixtitles{no} 
\appendix

\reftitle{References}


\externalbibliography{yes}
\bibliography{Coherent_eddies}





\end{document}